# Size-dependent Dielectric Permittivity of Perovskite Nanocrystals

*Jehyeok Ryu[1,3], Victor Krivenkov[2,3], Vitaly Goryashko[4], Yury Rakovich[1,2,3,5]\*, Alexey Y. Nikitin[1,5]\**

**Author Address**

[1] Donostia International Physics Center (DIPC), Donostia-San Sebastián 20018, Spain.

[2] Centro de Física de Materiales (CFM-MPC), Donostia - San Sebastián, 20018, Spain

[3] Polymers and Materials: Physics, Chemistry and Technology, Chemistry Faculty, University of the Basque Country (UPV/EHU), Donostia—San Sebastián 20018, Spain

[4] FREIA Laboratory, Department of Physics and Astronomy, Uppsala University, Ångströmlaboratoriet, 75120 Uppsala, Sweden

[5] IKERBASQUE, Basque Foundation for Science, Bilbao 48013, Spain

*Corresponding authors

yury.rakovich@ehu.eus, alexey@dipc.org

## Abstract

Perovskite nanocrystals (PNCs) are promising building blocks for quantum photonic devices. PNC's optical properties can be enhanced by integration with optical cavities/nanoantennas. Designing such structures requires accurate size-dependent dielectric permittivity of PNCs. However, current reports provide primarily ensemble-averaged values with limited access to the intrinsic response of individual PNCs. Here we suggest a methodology to reconstruct the size-dependent complex dielectric permittivity of $CsPbBr_3$ PNCs from the measured absorbance spectrum of colloidal solution. The PNC's permittivity is modeled as a sum of Voigt-profile oscillators, with the size-dependent transition energies governed by the exciton effective mass. Using a transmission electron microscopy-derived size distribution of the PNCs, the solution permittivity is obtained via Maxwell–Garnett effective medium approximation. This permittivity is used in a transfer-matrix method to simulate and fit the absorbance spectrum, from which the PNC's permittivity is reconstructed. The extracted spectral linewidth from the imaginary part of the permittivity (78.4 meV) is consistent with single-nanocrystal emission linewidths at room temperature. Finite-element simulations show enhanced absorption cross-section of a single PNC coupled to a nanoantenna, demonstrating applicability of the extracted permittivity. More generally, these findings provide a route to extract intrinsic permittivity of individual nanocrystals from absorbance measurements of their ensembles.

# 1. Introduction

PNCs have emerged as a versatile class of materials for optoelectronic and quantum photonic technologies. Their near-unity photoluminescence quantum yield[1–3], defect tolerance[4,5], and facile, low-cost solution based synthesis[1,3] make them attractive for light-emitting diodes[6,7], solar cells[8–10], lasers[11–13], and photodetectors[14]. In addition, their capability to operate as quantum emitters—featuring high single-photon purity, short radiative lifetimes, and partial optical coherence[15,16]—extends their potential to quantum communications.

To exploit these advantages in practical devices, PNCs are increasingly incorporated into dielectric cavities[17–20] or plasmonic nanoantennas[21–23] (metallic nanoparticles supporting localized surface plasmon polaritons (SPP) - electromagnetic fields coupled to oscillating charges-) where confined electromagnetic fields govern emission and absorption processes[20,24]. The optical performance of such hybrid systems depends on the complex dielectric permittivity, $\varepsilon$, of both the PNCs and their environment across spatial and photon energy, $E$, domains. Furthermore, the dielectric permittivity of PNCs, $\varepsilon_{PNC}$, varies with the size of the PNCs, L (the edge length of cube-shaped PNCs)[25–27]. However, to the best of our knowledge, most existing studies still extract ensemble-averaged dielectric permittivities[28–30]. For instance, size-dependent permittivities have been inferred by comparing nanocrystal ensembles with different average sizes for organic lead halide PNCs (MAPbX$_3$, X=Br, I)[28]. While this approach captures a trend with the mean size, it implicitly assumes that each ensemble is sufficiently monodisperse such that intra-ensembles size dispersion does not significantly contribute to the measured dielectric response. In practice, colloidal PNC solutions typically exhibit a finite size distribution, causing the extracted permittivity to represent a convolution over multiple nanocrystal sizes rather than a well-defined size resolved value. As a result, isolating intrinsic size-dependent dielectric properties would require near-ideal monodispersity across multiple ensembles, which is experimentally challenging to achieve.

Similarly, in weakly confined systems for CsPbBr$_3$ PNC, size effects are often assumed to be negligible, although the dielectric response remains effectively averaged over a finite size distribution[29]. This lack of information about the *L*-dependent dielectric permittivity of PNCs in the energy domain, $\varepsilon_{PNC}(E; L)$, hinders the precise design of PNCs-based quantum optics architectures. For instance, knowledge of $\varepsilon_{PNC}(E; L)$ enables quantitative modeling of light-matter interactions at the nanoscale involving PNCs embedded in structured electromagnetic environments, such as dielectric cavities or plasmonic nanoantennas. In such systems, accurate matching between the PNC size and confinement of geometry-dependent resonant modes of the surrounding nanophotonic structures can strongly enhance the optical performance of hybrid light-matter platforms.

Accurately determining the intrinsic $\varepsilon_{PNC}(E; L)$ is challenging for two main reasons. First, the absorption features of PNCs at room temperature exhibit non-Lorentzian spectral line shapes[31], reflecting pronounced inhomogeneous broadening due to inelastic phonon scattering. Second, the size dependence of excitonic transition energies deviates from the standard parabolic quantum-confinement relation, complicating the use of conventional size-dependent oscillator models. As a result, conventional Drude-Lorentz approaches[32] are generally insufficient to reproduce the experimentally observed absorption spectra of PNCs. To address these challenges, we develop an *L*-dependent dielectric permittivity model that unifies non-Lorentzian (Voigt) absorption line shapes and *L*-dependent energy levels of excitons within a single framework. We sequentially reconstruct the permittivity of a PNC solution ($\varepsilon_{sol}$) by averaging the dielectric permittivity of the solvent (toluene) and $\varepsilon_{PNC}(E; L)$ through Maxwell–Garnett effective-medium approximation[33]. Using $\varepsilon_{sol}$, we simulate an absorbance spectrum and fit it to the experimental one to reconstruct $\varepsilon_{PNC}(E; L)$. The resulting $\varepsilon_{PNC}(E; L)$ is further verified by comparing the spectral linewidth of the first peak of the imaginary part of the $\varepsilon_{PNC}(E; L)$ with the linewidth of single-PNC emission spectra to confirm their consistency. Finally, to demonstrate the practical applicability of the obtained $\varepsilon_{PNC}(E;L)$, we simulate the enhanced absorption cross-section of a PNC coupled to a plasmonic nanoresonator, revealing how the accurate size-dependent $\varepsilon_{PNC}(E;L)$ enables fine modeling of light–matter interactions at the nanoscale.

## 2. Results and Discussion

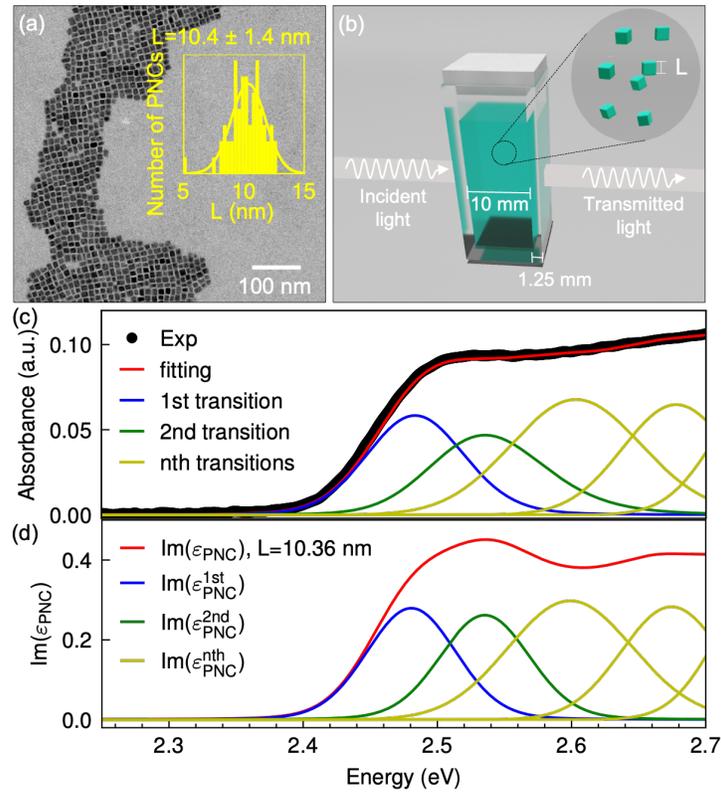

**Figure 1. Extraction of size-dependent dielectric permittivity of PNCs from solution (a)** Transmission electron microscopy image of the PNCs. The edge length distribution (inset) shows $L$=10.36 ($L_0$, the mean edge length) ± 1.42 ($\sigma_L$, the standard deviation of the edge length) nm. **(b)** Schematic illustration of light absorption by PNC colloidal solution composed of cubic nanocrystals. **(c)** Measured (black) and theoretically fitted (red) absorbance spectra of the PNC solution. Individual contributions to the absorbance from the 1st (blue), 2nd (green), and higher-order (nth, yellow) excitonic transitions are also shown. **(d)** Imaginary part of the extracted dielectric permittivity of PNCs at $L=L_0$, presented for the total response (red) and for each excitonic transition as above.

## 2.1. Modelling the dielectric permittivity of PNCs

The objective of this study is to obtain the permittivity of individual PNC, $\varepsilon_{PNC}(E;L)$ from the absorption of PNC ensembles dispersed in toluene, following the four steps: (i) synthesizing and characterizing the PNC solution, (ii) measuring the ultra-violet-to-visible (UV-Vis) absorbance spectrum of the PNC solution, (iii) modelling $\varepsilon_{PNC}(E;L)$ with fitting parameters and averaging it with the permittivity of toluene to reconstruct the $\varepsilon_{sol}$, and (iv) simulating the absorbance of the PNC solution using $\varepsilon_{sol}$ and fitting it to the experimental data to determine the fitting parameters of $\varepsilon_{PNC}(E;L)$.

First, we prepare a CsPbBr$_3$ PNC solution via ligand-assisted reprecipitation method[3,34] to characterize the PNCs in step (i). The solution consists of colloidal PNCs with a cube edge length, $L$, homogeneously dispersed in toluene with a molar concentration of ~23 nM. To reveal the size distribution of the colloidal PNCs in the solution, we perform imaging with transmission electron microscopy (TEM). **Figure 1a** presents the TEM image of the colloidal PNCs, confirming a cuboid morphology. An estimated average mean cube length is $L_0$ =10.36 nm and the standard deviation of the edge length, $\sigma_L$= 1.42 nm (see the inset in **Figure 1a**). Since the PNCs do not have an ideal monodisperse size-distribution, we assume that the optical properties of PNC ensembles in solution will be slightly deviated from those of individual PNCs. This result implies that the size-distribution of PNCs should be incorporated into the permittivity model.

To obtain the absorbance spectrum of the PNC solution, we load the PNC solution into a quartz cuvette with a wall thickness of 1.25 mm and an optical path length of 10 mm as schematically illustrated in **Figure 1b**. The unpolarized incident light with the intensity $I_0$ from a Xe lamp source is directed onto the sample, and the transmitted light intensity ($I$) is measured. The absorbance ($A$) is then calculated as $A = -log_{10}(I/I_0)$. The experimental absorbance spectrum is shown by the black curve in **Figure 1c**, revealing a continuous absorption onset around 2.4 eV. This continuum arises from multiple absorption spectral "lines" -bell-shaped resonance absorption spectral profiles-, which originate from transitions between different exciton energy levels. The first (lowest), second and higher-order ("n-th") absorption profiles are shown by the blue, green and yellow curves, respectively. The central energies of each absorption line also vary with $L$[35,36]. In addition, we hypothesize that each absorption line of individual PNCs have a Voigt profile shape[31] due to the thermal broadening at room temperature. Therefore, we employ the Brendel-Bormann (BB) model[37,38] that takes into account the Voigt profile shape (a Gaussian distribution of Lorentzian lineshapes) of each absorption line. We include $L$-dependent energies of the first five lower-energy absorption spectral lines to the BB model, so that $\varepsilon_{PNC}(E; L)$ has the following form:

$$\varepsilon_{PNC}(E; L) = \varepsilon_\infty + \sum_{n=1}^{5} \frac{iS_0 E_{p,n}}{\sqrt{8}\sigma_{ph,n}a}\left[\omega\left(\frac{a-E_n(L)}{\sqrt{2}\sigma_{ph,n}}\right) + \omega\left(\frac{a+E_n(L)}{\sqrt{2}\sigma_{ph,n}}\right)\right], \quad \textbf{(Equation 1)}$$

where $\varepsilon_\infty=7.3$, $\sigma_{ph,n}$ is the standard deviation to account for phonon-assisted inhomogeneous broadening of n-th absorption transition spectral line, $E_{p,n}$ is the Kane energy[35] -a characteristic energy of PNCs quantifying the oscillator strength of the n-th excitonic transition-, $S_0$ is an energy-scale prefactor, $E_n(L)$ is the energy corresponding to the n-th absorption transition, representing the gap between exciton's energy levels (see Supporting information), $\omega(z) = e^{-z^2} erfc(-iz)$ is Faddeeva function and $a = \frac{E}{\sqrt{2}}\left[\sqrt{1+\left(\frac{\gamma}{E}\right)^2}+1\right]^{1/2} + i\frac{E}{\sqrt{2}}\left[\sqrt{1+\left(\frac{\gamma}{E}\right)^2}-1\right]^{1/2}$, with $\gamma$ being the Lorentzian halfwidth at half maximum of the homogeneous broadening of the absorption line. Using this BB model, the absorption line shape can be properly described as a sum of multiple Voigt profiles. Additionally, the $L$-dependent energies of the first and second absorption transitions ($E_n(L)$, n=1,2) are obtained using the quasicubic size–dependent exciton energy level model[35]. Notice that the properly calculated absorption transition energies do not follow the parabolic quantum-confinement relation to the size, $\Delta E \sim \frac{\hbar^2 \pi^2}{2\mu L^2}$ –where $\hbar$ is the reduced Planck constant and $1/\mu = (1/m_e + 1/m_h)$ is the inverse of the effective mass of the exciton, with $m_e(m_h)$ being the mass of an electron (hole)– that neglects binding energy and the coupled bands of excitons[35]. Non-parabolic energy–size relationship can be properly taken into account by introducing a size-dependent effective mass of excitons, $\mu(L)$[35]. The $E_{n=1,2}(L)$ obtained from the quasicubic model are consistent with the experimentally-reported first and second excitonic energy levels[26] (Figure S1). Although experimentally-measured trends of the *L*-dependent energy levels with n≥3, to our knowledge, have not been reported (as they are hardly-distinguishable at room temperature[26]), we assume $E_n(L)$ (for n≥3) to follow the same trends as $E_1(L)$.

Next, we construct $\varepsilon_{sol}$ by combining the dielectric permittivity of toluene, $\varepsilon_{toluene}$, and $\varepsilon_{PNC}(E; L)$ via Maxwell-Garnett effective medium approximation. The nanocrystal size distribution with a gaussian profile, $f(L_0, \sigma_L) = \frac{f_0}{\sigma_L \sqrt{2\pi}} exp\left[-\frac{1}{2}\left(\frac{L-L_0}{\sigma_L}\right)^2\right]$ (where $f_0$ is the number concentration of the PNCs), is incorporated into $\varepsilon_{sol}$. This is done by averaging the Maxwell-Garnett relation over an ensemble of PNCs (see Supplementary Information for the details), so that $\varepsilon_{sol}=\varepsilon_{sol}(E; L_0, \sigma_L)$ is dependent upon $L_0$ and $\sigma_L$ that define any specific ensemble of PNCs. Finally, we simulate and fit the absorbance spectrum of the PNC solution as indicated by the red curve in **Figure 1c**, using $\varepsilon_{sol}(E; L_0, \sigma_L)$ in the transfer matrix method by varying unfixed parameters (see **Table 1** and Supporting Information for the details). The simulated absorbance spectrum perfectly matches the experimentally-measured one (black curve) across the full energy range for parameters indicated in Table S1. In a sharp contrast, applying the Drude-Lorentz model for PNCs, which doesn't account for Voigt-type broadening of spectral linewidths, fails to capture the feature near 2.4 eV (**Figure S3**), indicating that inhomogeneous broadening plays a significant role in the absorption transitions at room temperature. Consequently, the BB model combined with Maxwell-Garnett effective medium approximation enables us to consistently fit the experimental absorbance spectrum and thus reconstruct the PNC dielectric permittivity, within the model assumptions.

To visualize how individual absorption transitions contribute to the overall spectrum, in **Figure 1c** we plot the simulated absorbance spectra of the PNC solution for each n-th absorption transition with any other transitions being disregarded (blue, green, and yellow curves for the transitions with n=1, 2, n≥3, respectively). The spectra of individual absorption transitions overlap substantially due to their broad spectral linewidths ($\Delta E > 80$ meV) with respect to the distances between central peaks positions ($\Delta E < 80$ meV), giving rise to a nearly monotonous increasing absorbance spectrum (black and red curves in **Figure 1c**). **Figure 1d** demonstrates the imaginary part of the individual PNCs, $\text{Im}(\varepsilon_{PNC}(E; L_0))$ presumably proportional to the absorption by PNCs and thus responsible for the features in the absorption by the whole solution shown in **Figure 1c**. Similarly to the solution (red line in **Figure 1c**), $\text{Im}(\varepsilon_{PNC}(E; L_0))$ exhibits a continuous spectrum, with non-monotonic features emerging above $E \sim 2.54$ eV (red curve in **Figure 1d**). Such spectral shape arises from the superposition of five oscillators mimicking optical transitions from the ground state to individual excitonic levels in PNCs, as described by **Equation 1**. The contributions of the selected excitonic transitions are shown in **Figure 1d** (blue, green, yellow curves for n=1, n=2, n≥3, respectively). $\text{Im}(\varepsilon_{PNC}(E; L_0))$ spectrum exhibits a slightly different shape as compared to the one of the absorbance of the PNC solution, indicating that the latter is influenced by the convolution of the $\text{Im}(\varepsilon_{PNC}(E; L))$ with various values of *L*. As shown by the the red curve in **Figure 1d**, $\text{Im}(\varepsilon_{PNC}(E; L_0))$ already exhibits difficulty in distinguishing individual excitonic transition levels even for a PNC of a fixed size. Consequently, the lack of resolved peaks in the absorbance spectrum of the PNC solution (black and red curves in **Figure 1c**) originates predominantly from the intrinsic spectral overlap of broad spectral linewidths at room temperature, with size dispersion playing a secondary role. The $\varepsilon_{PNC}(E;L)$ obtained from this analysis is available (see Data Availability).

| n-th transitions | $E_{bulk,n}$ (eV) | $E_{p,n}$ (eV) | $\sigma_{ph,n}$ (meV) | $\gamma$ (meV) | $S_0$ (meV) | $\alpha_n$ | $\beta_n$ |
|---|---|---|---|---|---|---|---|
| 1st | 2.435 | 20.29 | 37 | 8 | 68 | 3 | 6.094 |
| 2nd | 2.435 | 19.44 | 37 | 8 | 65 | 6 | 5.014 |
| 3rd | 2.599 | 20.29 | 54 | 0 | 99 | 3 | 6.094 |
| 4th | 2.675 | 20.29 | 44 | 0 | 8 | 3 | 6.094 |
| 5th | 2.741 | 20.29 | 40 | 0 | 85 | 3 | 6.094 |

**Table 1. Fitting parameters used in the simulation.** The table lists seven key parameters for each *n*-th transition (*n*=1,…,5): the bulk exciton energy ($E_{bulk,n}$), Kane energy ($E_{p,n}$), phonon-assisted inhomogeneous broadening ($\sigma_{ph,n}$), intrinsic homogeneous broadening ($\gamma$), an energy scale factor ($S_0$), and empirical constants ($\alpha_n, \beta_n$). The size-dependent exciton effective mass $\mu_n$ is calculated as: $\mu_n = \dfrac{3m_0\left(\sqrt{E_{bulk,n}^2+\frac{2E_{p,n}}{m_0}\left(\frac{\hbar\pi}{L}\right)^2}+\sqrt{E_{bulk,n}^2+\frac{4E_{p,n}}{m_0}\left(\frac{\hbar\pi}{L}\right)^2}\right)}{4E_{p,n}}$. The size-dependent exciton energy levels, $E_n(L)$, are defined by: $E_n(L) = E_{bulk,n} + \dfrac{\alpha_n \pi^2 \hbar^2}{2\mu_n L^2} - B_x\sqrt{1+\left(\beta_n\frac{a_x}{L}\right)^2}$. Sequentially, $E_n(L), S_0, \sigma_{ph,n}, \gamma$ are used to compute $\varepsilon_{PNC}(E; L)$. Finally, $\varepsilon_{sol}(E; L_0, \sigma_L)$ is obtained via Maxwell-Garnett effective medium approximation (see the details in the Supporting Information).

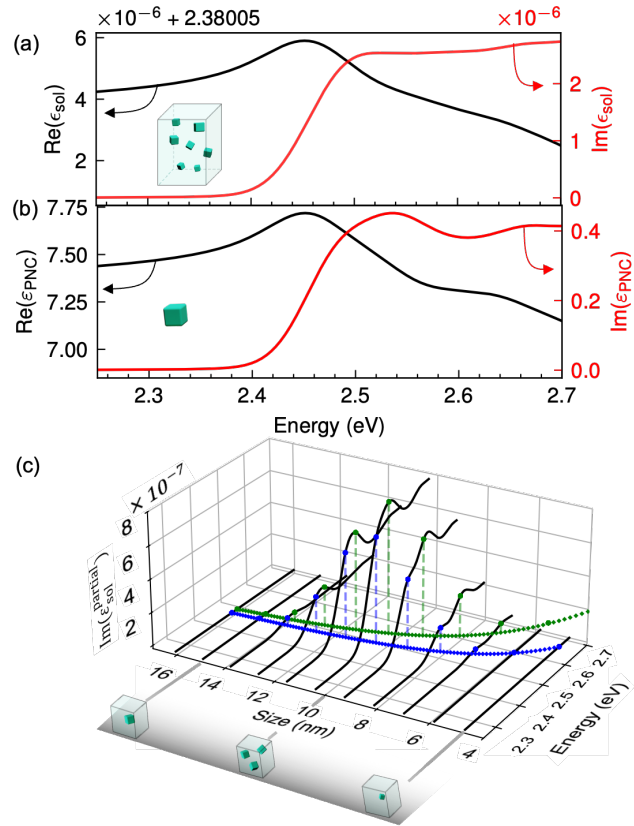

**Figure 2. Dielectric permittivity of PNCs and their colloidal solution. (a,b)** Real (black, left-axis) and imaginary (red, right-axis) parts of the dielectric permittivity of the PNC solution, $\varepsilon_{sol}(E; L_0, \sigma_L)$, panel (a) and of individual PNCs with edge length $L = L_0$, $\varepsilon_{PNC}(E; L_0)$, panel (b). **(c)** Decomposition of $\varepsilon_{sol}(E; L_0, \sigma_L)$ into contributions from different PNC sizes. Each gray curve represents the dielectric permittivity of a hypothetical solution containing only partial fraction of PNCs of a given size $L$, $\varepsilon_{sol}^{partial}(E; L, \Delta L)$, weighted by their volume fraction according to the size distribution in the inset of Figure 1b. The excitonic energy levels of the first and second transitions from the quasicubic model of the exciton energy levels are overlaid on the curves.

## 2.2. Effects of size distribution on dielectric permittivity

In order to elucidate how the optical response of a PNC solution relates to that of individual PNCs, we compare $\varepsilon_{sol}(E; L_0, \sigma_L)$ and $\varepsilon_{PNC}(E; L_0)$ for the PNC size of $L = L_0$. The real and imaginary parts of the extracted dielectric permittivities for the solution and for the individual PNCs are shown in **Figure 2a** and **Figure 2b**, respectively. A pronounced difference in magnitude between $\varepsilon_{sol}(E; L_0, \sigma_L)$ and $\varepsilon_{PNC}(E; L_0)$ is observed. In particular, near the first exciton feature seen at ~2.5 eV, the real (imaginary) part of $\varepsilon_{sol}(E; L_0, \sigma_L)$ reaches the value of approximately ~2.38 (~$2 \times 10^{-6}$), whereas that of $\varepsilon_{PNC}(E; L_0)$ attains much larger values ~7.5 (~0.4), respectively. The difference between $\varepsilon_{sol}(E; L_0, \sigma_L)$ and $\varepsilon_{toluene}$ =2.38 is of order of $10^{-6}$, while the difference between $\varepsilon_{PNC}(E; L_0)$ and $\varepsilon_{\infty}$=7.3 is much larger (order of $10^{-1}$). Despite this strong contrast in magnitude, the shape of the spectra for real and imaginary parts of $\varepsilon_{sol}(E; L_0, \sigma_L)$ closely follows the one of $\varepsilon_{PNC}(E; L_0)$, indicating the intrinsic optical response of the PNCs is retained in the solution, but substantially attenuated. This suppression in magnitude of $\varepsilon_{sol}(E; L_0, \sigma_L)$ originates from the small PNC's volume fraction in the solution. Specifically, the volume fraction of PNCs relative to the solvent is estimated to be ~ $1.8 \times 10^{-5}$, corresponding to a PNC molar concentration of 23 nM obtained from ICP–AES measurement.

While the overall spectral features are preserved, $\varepsilon_{PNC}(E; L_0)$ shows slight difference in the detailed spectral line shapes compared to $\varepsilon_{sol}(E; L_0, \sigma_L)$. We attribute this difference to the fact that $\varepsilon_{sol}(E; L_0, \sigma_L)$ arises from the superposition of contributions from PNCs with different sizes. To support this interpretation, we consider the optical response of isolated sub-ensembles of PNCs with a narrow range around a given *L*. Specifically, we define the imaginary part of the partial dielectric permittivities of PNC solutions, Im($\varepsilon_{sol}^{partial}(E; L, \Delta L)$), which is calculated by including only PNCs with sizes in the range $L \pm \Delta L/2$ in the solution, where $\Delta L$ ~ 0.11 nm denotes the size bin width used for the integration (see Equation S5 in the Supporting Information for details). As shown in **Figure 2c**, Im($\varepsilon_{sol}^{partial}(E; L, \Delta L)$) exhibits its maximum magnitude around *L*~10 nm and decreases away from this value, consistently with the PNC size distribution shown in the inset of **Figure 1a**. To further elucidate how size-dependent excitonic transitions shape the spectrum of Im($\varepsilon_{sol}^{partial}(E; L, \Delta L)$), we overlay the size-dependent first and second excitonic energies obtained from the quasicubic exciton energy level model ($E_n(L)$, n=1,2) in the energy-size plane of **Figure 2c**. The $E_1(L)$ and $E_2(L)$ are indicated by the blue and green vertical dashed lines ending up at the dots on the corresponding Im($\varepsilon_{sol}^{partial}(E; L, \Delta L)$) grey curves. For smaller nanocrystals with $L < 8\ nm$, the two excitonic transition energy levels are well separated, with $E_2(L) - E_1(L) > 0.1\ eV$. In contrast, for larger nanocrystals with $L > 10\ nm$, the energy separation decreases below $0.05\ eV$, leading to a substantial spectral overlap and merging into a single peak. This observation indicates that the single peak observed in the Im($\varepsilon_{sol}^{partial}(E; L, \Delta L)$) for large nanocrystals ($L > 10\ nm$) originates from overlapping excitonic transitions rather than from a single exciton transition. Similar exciton spectral merging behavior has been reported for large CsPbBr$_3$ PNCs thin films at 15 K[25].

Overall, the $\varepsilon_{sol}(E; L_0, \sigma_L)$ incorporates the effects of both the suppression of the intrinsic PNC optical response due to the small PNC volume fraction in the PNC solution and the convolution of size-dependent excitonic features, where smaller PNCs retain distinct excitonic peaks while larger PNCs exhibit merged transitions due to the spectral overlap.

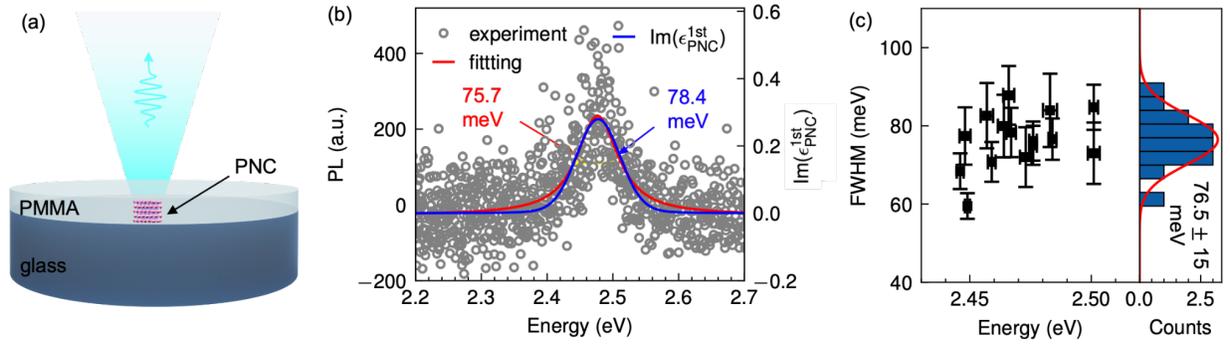

**Figure 3. Comparison between the linewidths of the first excitonic transition in the complex dielectric permittivity and the emission spectrum of a single PNCs. (a)** Schematic illustration of the PNC sample embedded in a PMMA layer on a glass substrate for emission measurements. **(b)** Experimental emission spectrum of a single PNC (gray open circles) and its Voigt–profile fit (red curve), yielding a FWHM of ~ 75.7 meV. The imaginary part of the dielectric permittivity corresponding to the first exciton transition, $\text{Im}(\varepsilon_{PNC}^{1st}(E;L))$, is shown as the blue curve, exhibiting a FWHM ~ 78.4 meV, in close agreement with the experimental value. **(c)** Statistical distribution of the FWHM values obtained from emission spectra of 15 PNCs, showing a mean FWHM of 76.5 ± 15 meV, consistent with the theoretical result.

## 2.3. Imaginary part of dielectric permittivity and single PNC linewidth

To corroborate the obtained fitting results for $\varepsilon_{PNC}(E;L)$, we perform photoluminescence (PL) measurement of single PNCs and compare the measured PL linewidth profile with that of the imaginary part of the dielectric permittivity of PNCs, accounting for the first exciton energy level ($\text{Im}(\varepsilon_{PNC}^{1st}(E;L))$, $n = 1$ from **Equation 1**). Since PL emission arises from radiative recombination of an exciton from the first (lowest) excitonic excited state to the ground state, following rapid non-radiative relaxations of excitons from higher excitonic states in PNCs[39], the linewidth of the first exciton transition provides a relevant reference. **Figure 3a** schematically illustrates the PL measurement. Isolated single PNCs are embedded into a PMMA layer on a glass substrate, and PL spectra are collected from individual nanocrystals. **Figure 3b** shows a representative PL spectrum (grey open circles) and its Voigt-profile fit (red curve), yielding a full width at half maximum (FWHM) ~75.7 meV. This linewidth is comparable to the FWHM value of ~78.4 meV obtained from the $\text{Im}(\varepsilon_{PNC}^{1st}(E;L))$ spectrum (blue curve in Figure 3b). We repeat the single-particle PL measurement on fourteen nanocrystals to obtain statistical linewidth data (**Figure 3c**), which yield an average FWHM of ~76.5 meV, again comparable to the modeled value. The consistent linewidths obtained from the PL experiments and the $\text{Im}(\varepsilon_{PNC}^{1st}(E;L))$ indicate that the modeled broadening realistically reflects the PL spectral shape of the excitonic transition of individual PNCs. Therefore, our dielectric permittivity model supports the essential optical behavior of single colloidal PNCs at room temperature.

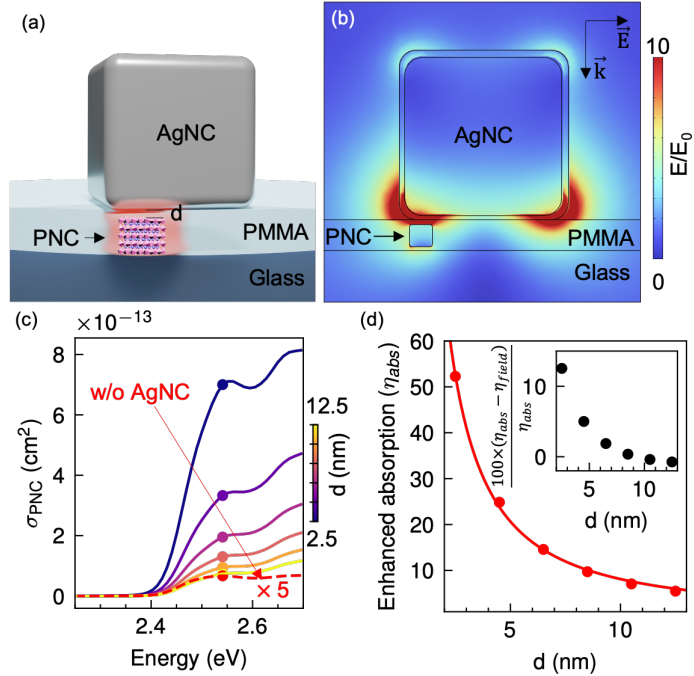

**Figure 4. Finite element method simulation of the absorption cross–section of a PNC and its enhancement through coupling to a plasmonic cavity.** **(a)** Schematic illustration of the simulated PNC–plasmonic cavity system. The PNC is embedded in a PMMA layer on a glass substrate, and a silver nanocube (AgNC) with an edge length of 75 nm is placed on top of the PMMA layer. The vertical separation between the bottom facet of the AgNC and the top of the PNC is defined as the gap distance *d*. **(b)** Electric-field distribution of the structure, showing strong near–field localization ("hot spots") at the two bottom corners of the AgNC. **(c)** Calculated absorption cross-section of a PNC without coupling (red dashed curve) and with plasmonic coupling for different gap distances. For clarity, the uncoupled spectrum is multiplied by a factor of five. **(d)** Dependence of the enhanced absorption, $\eta_{abs}$, on the gap distance. The red curve shows a fit to the $\eta_{abs}$ as a function of *d*. (inset) Relative difference between the absorption enhancement factor, $\eta_{abs}$, and the local field intensity enhancement factor, $\eta_{field}$. The percentage difference is defined as $100 \times (\eta_{abs} - \eta_{field})/\eta_{abs}$.

## 2.4. Application of the extracted dielectric permittivity for a plasmonic cavity design

We will now demonstrate how our reconstructed dielectric permittivity of PNCs can be used to accurately analyse light-matter interactions in realistic hybrid systems. In particular, we compute the absorption cross-section of a PNC coupled to a plasmonic nanoresonator - a silver nanocube (AgNC) – using the PNC-size-dependent model of $\varepsilon_{PNC}(E; L)$ in finite-element-method (FEM) electromagnetic simulations. The nanoresonator is illuminated by an external laser field so that the electric field re-distributes in space creating regions with very high intensity of the near-field – usually at discontinuities like edges and corners. If a PNC is placed in the region of the high-intensity near-field, it will exhibit a large absorption cross-section. In practical applications, this implies that the same excitation efficiency of the PNC can be reached with a much lower incident laser intensity, thus, relaxing the power requirements on the pump laser. At the same time, in order to prevent the laser damage of the PNC, we need to compute precisely the absorption cross-section enhancement factor. Hence, the accurate reconstruction of $\varepsilon_{PNC}(E; L)$ performed earlier is a must. In the following analysis, we restrict ourselves to the linear optical regime, where the dielectric response of the PNC is assumed to be independent of excitation intensity and absorption saturation effects are neglected.

**Figure 4a** shows the geometry of the electromagnetic model: a PNC is embedded into a PMMA layer lying on a glass substrate and positioned in the proximity of a silver nanocube (AgNC) with an edge length of 75 nm. The gap (*d*) between the bottom surface of the AgNC and the top surface of the PNC is scanned in simulations. **Figure 4b** presents the spatial distribution of the electric-field enhancement factor, $|E|/|E_0|$, for *d* = 2.5 nm, where $E_0$ is the electric field of the incoming plane wave polarized along the x-axis. The incident wave excites localized SPP in the cube. A strong field confinement of the SPPs electric field to the nanocube's corners is observed, indicating that the PNC experiences a highly nonuniform field distribution over its volume, which justifies the need for a volumetric absorption evaluation rather than a single-point field analysis.

We calculate the effective absorption cross-section of the PNC (absorbed power by the PNC, $p_{PNC} = \frac{E}{2\hbar} \int_V \varepsilon_{PNC}(E;L)|E(r)|^2 dV$, normalized to the incident radiation power flow $I_0$) in the absence of AgNC, $\sigma_{PNC}^0 = \frac{p_{PNC}^0}{I_0}$, and in the presence of AgNC, $\sigma_{PNC} = \frac{p_{PNC}}{I_0}$, respectively (**Figure 4c**). Note that the absorbed power $p_{PNC}$ is assumed to scale linearly with the local electric field intensity $|E(r)|^2$, thereby neglecting the absorption saturation effects. Consequently, this approach is valid only in the weak excitation regime, where $I_0$ remains well below the saturation intensity. In the absence of AgNC, $\sigma_{PNC}^0$ (red dashed curve in Figure 4c), scaled by a factor of five for clarity, exhibits a peak (red dot in **Figure 4c**) with the maximal magnitude of approximately $1.34 \times 10^{-14}$ cm$^2$ at around 2.54 eV – the same order of magnitude as in previously reported results[29,40]. The agreement in both the absorption cross-section and the linewidth (**Figure 3**) supports the quantitative reliability of the extracted dielectric permittivity. This result implies that the absorption cross-section of individual PNCs can be quantitatively obtained by different particle sizes. For example, this information can be useful in modeling excitonic dynamics at the single-PNC level, where the optical pumping rate is given by the product of the absorption cross-section and the incident power density.

Furthermore, we compare the absorption cross-section enhancement factor, $\eta_{abs} = \frac{\sigma_{PNC}}{\sigma_{PNC}^0}$ (red dots in **Figure 4d and Figure S4**), with the commonly used electric-field intensity enhancement, $\eta_{field} = \frac{|E|^2}{|E_0|^2}$ (grey dots in **Figure S4**) as a function of the gap distance *d*. Both quantities increase as the gap distance decreases, reflecting the strengthening electromagnetic coupling between the PNC and the AgNC. At larger separations, where the near field is approximately uniform across the PNC volume, $\eta_{abs}$ and $\eta_{field}$ show good agreement. In contrast, at smaller gap distance, a modest deviation (~10%) becomes apparent (inset of Figure 4d), which arises from stronger spatial variations of the near field across the finite PNC volume that are captured by $\eta_{abs}$ but not by $\eta_{field}$.

In summary, incorporating the extracted PNC dielectric permittivity into FEM simulations allows a direct and realistic evaluation of absorption cross-section of PNCs in electromagnetically-coupled systems. This approach captures both spatial and spectral field variations, as well as the intrinsic optical response of the PNCs, and provides a robust framework for designing cavity and antenna geometries optimized for perovskite-based single-photon sources and nanophotonic devices.

## 3. Conclusion

In this study, we establish a protocol to extract the complex, size-dependent dielectric permittivity of CsPbBr$_3$ perovskite nanocrystals directly from the absorbance spectrum of a colloidal solution. By combining the Brendel-Bormann model with the Maxwell-Garnett effective medium approximation, we accurately reproduce the measured absorbance features of a PNC solution and obtain the size–dependent complex dielectric permittivity of individual nanocrystals. The extracted dielectric permittivity yields exciton linewidths consistent with single-particle photoluminescence measurements conducted here.

Furthermore, incorporation of the extracted dielectric permittivity into finite-element electromagnetic simulations enables a quantitative evaluation of the absorption cross-section of PNCs, being consistent with the reported values, thus validating our approach. Importantly, this method provides a detailed assessment of the absorption enhancement in electromagnetically-coupled geometries, such as single PNCs positioned in the near-field of a metallic nanoresonator. This assessment goes beyond conventional estimates based on electromagnetic field enhancement at a fixed point (e.g. $\frac{|E|^2}{|E_0|^2}$ evaluated at the center of a PNC) by accounting for the spatial variation of the near field over the entire nanocrystal volume.

As a result, our framework bridges optical spectroscopy and electromagnetic modeling, providing a practical route to parameterize perovskite nanocrystals through their size-dependent dielectric permittivity for device-scale simulations. Our findings can be useful for various practical implications, such as determining the concentration of PNC solutions. From a more general perspective, our methodology can be extended to other halide PNCs as well as to other, conventional (II-VI) colloidal quantum dots, enabling predictive design of nanophotonic architectures for light emission, detection, and quantum optical applications.

## 4. Experimental Section

**Synthesis of CsPbBr$_3$ Perovskite Nanocrystals**

CsPbBr$_3$ PNCs were prepared via a ligand-assisted reprecipitation (LARP) approach conducted under ambient conditions. In a typical synthesis, 17.9 mg of cesium bromide (CsBr, 99.999%, Sigma-Aldrich) was dissolved in 2 mL of anhydrous dimethylformamide (DMF, Sigma-Aldrich) and sonicated for approximately 2 h until a clear solution was obtained. Subsequently, 30.8 mg of lead bromide (PbBr$_2$, 99.999%, Sigma-Aldrich) was introduced into the same vial and further sonicated for 10 min to ensure complete dissolution.

To this precursor mixture, 200 µL of oleic acid (technical grade, 90%) and 100 µL of oleylamine (> 98%, primary amine, Sigma-Aldrich) were added while stirring vigorously for 10 min. A 40 µL aliquot of the resulting stock solution was then swiftly injected into 2 mL of anhydrous toluene under continuous stirring. The dispersion immediately turned greenish-yellow, confirming the formation of CsPbBr$_3$ PNCs. For surface passivation, 2.5 µL of a phenethylammonium bromide (PEABr, >98%, Sigma-Aldrich) solution in DMF (200 mg mL$^{-1}$) was added, followed by an additional 2 h of stirring. The colloidal suspension was allowed to age for 3 days and subsequently centrifuged at 7500 rpm for 5 min. The clear supernatant was collected and diluted three– to four–fold to achieve an absorbance of approximately 0.2 at the first excitonic absorption peak. This value was empirically chosen as a convenient concentration reference, such that a subsequent dilution yielded sparsely distributed, isolated single QDs upon spin coating.

**Preparation of a Film Sample for Single PNC Measurements**

A Film for single–particle studies was fabricated by diluting 5 µL of the as–prepared PNC solution in 395 µL of a 1 wt% polymethyl methacrylate (PMMA) solution in anhydrous toluene. A 400 µL aliquot of this diluted mixture was drop–cast onto a clean glass coverslip, ensuring full surface coverage, and spin–coated under ambient conditions at 1000 rpm for 1 min. The resulting film was immediately suitable for single–particle optical measurements.

**Transfer matrix simulations**

The absorbance spectra of the PNC solution were simulated using the transfer–matrix method implemented through a Python–based open–source library[41].

**Corresponding Author**

Yury Rakovich (yury.rakovich@ehu.eus), Alexey Y. Nikitin (alexey@dipc.org)

**Acknowledgement**

The study was funded by the Department of Science, Universities and Innovation of the Basque Government (grants no. IT1526-22, PIBA_2024_1_0011, and PIBA-2023-1-0007) and the IKUR Strategy; by the Spanish Ministry of Science and Innovation (grants no. PID2022-141017OB-I00, TED2021-129457B-I00, PID2023-146442NB-I00, PID2023-147676NB-I00). VG acknowledges funding from the Swedish Research Council (Vetenskaprådet, 2022-03983).

**Data Availability Statement**

The data that support the findings of this study are available from the corresponding author upon reasonable request. The Python source code used to extract and export the size-dependent complex dielectric permittivity is freely available at https://doi.org/10.5281/zenodo.18294506

# Reference

1. Protesescu, L. *et al.* Nanocrystals of Cesium Lead Halide Perovskites (CsPbX3, X = Cl, Br, and I): Novel Optoelectronic Materials Showing Bright Emission with Wide Color Gamut. *Nano Lett.* **15**, 3692–3696 (2015).

2. Shamsi, J., Urban, A. S., Imran, M., De Trizio, L. & Manna, L. Metal Halide Perovskite Nanocrystals: Synthesis, Post-Synthesis Modifications, and Their Optical Properties. *Chem. Rev.* **119**, 3296–3348 (2019).

3. Li, X. *et al.* CsPbX3 Quantum Dots for Lighting and Displays: Room-Temperature Synthesis, Photoluminescence Superiorities, Underlying Origins and White Light-Emitting Diodes. *Advanced Functional Materials* **26**, 2435–2445 (2016).

4. Akkerman, Q. A., Rainò, G., Kovalenko, M. V. & Manna, L. Genesis, challenges and opportunities for colloidal lead halide perovskite nanocrystals. *Nature Mater* **17**, 394–405 (2018).

5. Kang, J. & Wang, L.-W. High Defect Tolerance in Lead Halide Perovskite CsPbBr3. *J. Phys. Chem. Lett.* **8**, 489–493 (2017).

6. Chen, W. *et al.* Highly bright and stable single-crystal perovskite light-emitting diodes. *Nat. Photon.* **17**, 401–407 (2023).

7. Liu, Y. *et al.* Bright and Stable Light-Emitting Diodes Based on Perovskite Quantum Dots in Perovskite Matrix. *J. Am. Chem. Soc.* **143**, 15606–15615 (2021).

8. Park, S. M. *et al.* Low-loss contacts on textured substrates for inverted perovskite solar cells. *Nature* **624**, 289–294 (2023).

9. Park, S. M. *et al.* Engineering ligand reactivity enables high-temperature operation of stable perovskite solar cells. *Science* **381**, 209–215 (2023).

10. Lin, R. *et al.* All-perovskite tandem solar cells with improved grain surface passivation. *Nature* **603**, 73–78 (2022).

11. Sun, W. *et al.* Lead halide perovskite vortex microlasers. *Nat Commun* **11**, 4862 (2020).

12. Zhang, Q., Shang, Q., Su, R., Do, T. T. H. & Xiong, Q. Halide Perovskite Semiconductor Lasers: Materials, Cavity Design, and Low Threshold. *Nano Lett.* **21**, 1903–1914 (2021).

13. Zhu, H. *et al.* Lead halide perovskite nanowire lasers with low lasing thresholds and high quality factors. *Nature Mater* **14**, 636–642 (2015).

14. He, Y. *et al.* High spectral resolution of gamma-rays at room temperature by perovskite CsPbBr3 single crystals. *Nat Commun* **9**, 1609 (2018).

15. Kaplan, A. E. K. *et al.* Hong–Ou–Mandel interference in colloidal CsPbBr3 perovskite nanocrystals. *Nat. Photon.* **17**, 775–780 (2023).

16. Utzat, H. *et al.* Coherent single-photon emission from colloidal lead halide perovskite quantum dots. *Science* **363**, 1068–1072 (2019).

17. Farrow, T. *et al.* Ultranarrow Line Width Room-Temperature Single-Photon Source from Perovskite Quantum Dot Embedded in Optical Microcavity. *Nano Lett.* **23**, 10667–10673 (2023).

18. He, R. *et al.* Interplay of Purcell effect and extraction efficiency in CsPbBr3 quantum dots coupled to Mie resonators. *Nanoscale* **15**, 1652–1660 (2023).

19. Jun, S. *et al.* Ultrafast and Bright Quantum Emitters from the Cavity-Coupled Single Perovskite Nanocrystals. *ACS Nano* **18**, 1396–1403 (2024).

20. Purkayastha, P. *et al.* Purcell Enhanced Emission and Saturable Absorption of Cavity-Coupled CsPbBr3 Quantum Dots. *ACS Photonics* **11**, 1638–1644 (2024).

21. Olejniczak, A. *et al.* On-demand reversible switching of the emission mode of individual semiconductor quantum emitters using plasmonic metasurfaces. *APL Photonics* **9**, 016107 (2024).

22. Olejniczak, A., Rakovich, Y. & Krivenkov, V. Advancements and challenges in plasmon-exciton quantum emitters based on colloidal quantum dots and perovskite nanocrystals. *Mater. Quantum. Technol.* **4**, 032001 (2024).

23. Hsieh, Y.-H. *et al.* Perovskite Quantum Dot Lasing in a Gap-Plasmon Nanocavity with Ultralow Threshold. *ACS Nano* **14**, 11670–11676 (2020).

24. Novotny, L. & Hecht, B. *Principles of Nano-Optics*. (Cambridge University Press, Cambridge, 2012). doi:10.1017/CBO9780511794193.

25. Boehme, S. C. *et al.* Single-photon superabsorption in CsPbBr3 perovskite quantum dots. *Nat. Photon.* 1–7 (2025) doi:10.1038/s41566-025-01684-3.

26. Krieg, F. *et al.* Monodisperse Long-Chain Sulfobetaine-Capped CsPbBr3 Nanocrystals and Their Superfluorescent Assemblies. *ACS Cent. Sci.* **7**, 135–144 (2021).

27. Hou, L., Tamarat, P. & Lounis, B. Revealing the Exciton Fine Structure in Lead Halide Perovskite Nanocrystals. *Nanomaterials* **11**, 1058 (2021).

28. Rubino, A., Lozano, G., E. Calvo, M. & Míguez, H. Determination of the optical constants of ligand-free organic lead halide perovskite quantum dots. *Nanoscale* **15**, 2553–2560 (2023).

29. Park, S.-H. *et al.* Complex Refractive Index Spectrum of CsPbBr3 Nanocrystals via the Effective Medium Approximation. *Nanomaterials* **15**, 181 (2025).27

30. Rodríguez Ortiz, F. A. *et al.* The Anisotropic Complex Dielectric Function of CsPbBr3 Perovskite Nanorods Obtained via an Iterative Matrix Inversion Method. *J. Phys. Chem. C* **127**, 14812–14821 (2023).

31. Xu, C., Vollbrecht, J. & Clausing, R. Analytical model for optical permittivity in direct bandgap semiconductors with Gaussian distributed bandgap energies. *Opt. Lett., OL* **50**, 371–374 (2025).

32. Vasilevskiy, M. I. & Anda, E. V. Effective dielectric response of semiconductor composites. *Phys. Rev. B* **54**, 5844–5851 (1996).

33. Markel, V. A. Introduction to the Maxwell Garnett approximation: tutorial. *J. Opt. Soc. Am. A* **33**, 1244 (2016).

34. Ryu, J. *et al.* Nickel Doping Unlocks Ambient-Condition Photostability in Individual Cesium Lead Bromide Perovskite Quantum Dots. *Nano Lett.* **25**, 16630–16636 (2025).

35. Sercel, P. C., Lyons, J. L., Bernstein, N. & Efros, A. L. Quasicubic model for metal halide perovskite nanocrystals. *The Journal of Chemical Physics* **151**, 234106 (2019).

36. Sercel, P. C. *et al.* Exciton Fine Structure in Perovskite Nanocrystals. *Nano Lett.* **19**, 4068–4077 (2019).

37. Brendel, R. & Bormann, D. An infrared dielectric function model for amorphous solids. *J. Appl. Phys.* **71**, 1–6 (1992).

38. Orosco, J. & Coimbra, C. F. M. Optical response of thin amorphous films to infrared radiation. *Phys. Rev. B* **97**, 094301 (2018).

39. Ryu, J., Krivenkov, V., Olejniczak, A., Nikitin, A. Y. & Rakovich, Y. Perovskite nanocrystals as emerging single-photon emitters: Progress, challenges, and opportunities. *Appl. Phys. Rev.* **12**, 041323 (2025).

40. Ravi, V. K., Swarnkar, A., Chakraborty, R. & Nag, A. Excellent green but less impressive blue luminescence from CsPbBr3 perovskite nanocubes and nanoplatelets. *Nanotechnology* **27**, 325708 (2016).

41. Byrnes, S. J. Multilayer optical calculations. *arXiv*: 1603.02720 (2021).